\pdfoutput=1
\documentclass[12pt]{mn2e}
\usepackage{graphicx}
\usepackage{epsf}
\usepackage{lscape}
\usepackage{longtable}
\usepackage{rotating}
\usepackage{color}

\newcommand{\aap}{A\&A}

\newcommand{\apj}{ApJ}
\newcommand{\apjl}{ApJ}

\newcommand{\mnras}{MNRAS}

\definecolor{grey}{rgb}{0.7,0.7,0.7}

\begin{document}
\topmargin -0.5in 

\title[Interpreting observed UV continuum slopes]{Interpreting the observed UV continuum slopes of high-redshift galaxies}

\author[Stephen M. Wilkins et al.]  
{
Stephen M. Wilkins$^{1}$\thanks{E-mail: stephen.wilkins@physics.ox.ac.uk}, Andrew Bunker$^{1}$, William Coulton$^{1}$, Rupert Croft$^{1,2}$,\newauthor Tiziana Di Matteo$^{1,2}$, Nishikanta Khandai$^{2,3}$, Yu Feng$^{2}$\\
$^1$\,University of Oxford, Department of Physics, Denys Wilkinson Building, Keble Road, OX1 3RH, U.K. \\
$^2$\,McWilliams Center for Cosmology, Carnegie Mellon University, 5000 Forbes Avenue, Pittsburgh, PA 15213, U.S.A.\\
$^3$\,Brookhaven National Laboratory, Department of Physics, Upton, NY 11973, U.S.A.
}

\maketitle 

\begin{abstract}
The observed UV continuum slope of star forming galaxies is strongly affected by the presence of dust. Its observation is then a potentially valuable diagnostic of dust attenuation, particularly at high-redshift where other diagnostics are currently inaccesible. Interpreting the observed UV continuum slope in the context of dust attenuation is often achieved assuming the empirically calibrated Meurer et al. (1999) relation. Implicit in this relation is the assumption of an intrinsic UV continuum slope ($\beta=-2.23$). However, results from numerical simulations suggest that the intrinsic UV continuum slopes of high-redshift star forming galaxies are bluer than this, and moreover vary with redshift. Using values of the intrinsic slope predicted by numerical models of galaxy formation combined with a Calzetti et al. (2000) reddening law we infer UV attenuations ($A_{1500}$) $0.35-0.5\,{\rm mag}$ ($A_{V}$: $0.14-0.2\,{\rm mag}$ assuming Calzetti et al. 2000) greater than simply assuming the Meurer relation. This has significant implications for the inferred amount of dust attenuation at very-high ($z\approx 7$) redshift given current observational constraints on $\beta$, combined with the Meurer relation, suggest dust attenuation to be virtually zero in all but the most luminous systems.
\end{abstract} 

\begin{keywords}  
galaxies: evolution –- galaxies: formation –- galaxies: starburst –- galaxies: high-redshift –- ultraviolet: galaxies
\end{keywords} 

\section{Introduction}

The intrinsic UV luminosity of a galaxy is a frequently used diagnostic of star formation (e.g. Wilkins et al. 2012b). However, UV emission is easily attenuated by dust. Determining the level of dust attenuation is then critical to accurately derive intrinsic UV luminosities and thus star formation rates. At high-redshift, far-IR observations, which probe dust-reprocessed UV emission (and thus obscured star formation), are currently inaccesible for all but the most luminous galaxies.

Using observations taken with Wide Field Camera 3 (WFC3) on the {\em Hubble Space Telescope} means it is now routinely possible to identify star forming galaxies at $z=7-8$ (e.g. Oesch et al. 2010, Bouwens et al. 2010b, Bunker et al. 2010, Wilkins et al. 2010, Finkelstein et al. 2010, McLure et al. 2010, Wilkins et al. 2011a, Lorenzoni et al. 2011, Bouwens et al. 2011, McLure et al. 2011, Finkelstein et al. 2012b, Lorenzoni et al. 2012) and even to $z\sim 10$ (e.g. Oesch et al. 2012). These deep near-IR observations also allow us to probe the rest-frame UV continuum (with at least 2 filters unattenuated by the Lyman-$\alpha$ break) at $z=4-7$ allowing the measurement of the UV continuum slope $\beta$ (defined\footnote{Alternatively $f_{\nu}\propto \lambda^{\beta+2}$.} such that $f_{\lambda}\propto\lambda^{\beta}$). 

The observed UV continuum slope has been demonstrated to provide a reasonable diagnostic of moderate dust attenuation in local star forming galaxies (e.g. Meurer et al. 1999, hereafter M99). By virtue of being UV selected, the UV continuum slopes of high-redshift galaxies are relatively easy to measure, even for the faintest sources, as they require observations in only a single additional filter relative to the three required to robustly identify high-redshift galaxies. Consequently the UV continuum slope is one of the few accessible diagnostics of galaxies that are available at very high-redshift and has been subject to significant interest (e.g. Stanway et al. 2005, Bouwens et al. 2009, Bunker et al. 2010, Bouwens et al. 2010a, Wilkins et al. 2011b, Dunlop et al. 2012a, Bouwens et al. 2012, Finkelstein et al. 2012a, Rogers, McLure \& Dunlop 2012, Dunlop et al. 2012b).

The observed UV continuum slope (of star forming galaxies) can be interpreted, in the context of dust, with knowledge of both the intrinsic UV continuum slope (of star forming galaxies) and the reddening law. A shortcut is to use a relation empirically derived using observations of the UV continuum slope and far-IR emission such as that proposed by M99 (though this implicitly assumes an intrinsic UV continuum slope). However, the M99 relation is calibrated at low-redshift, where the properties of star forming galaxies, and in particular the {\em intrinsic} UV continuum slopes may be significantly different from the systems we can now study at high-redshift. Indeed recent predictions from the {\sc galform} semi-analytical model of galaxy formation (see Wilkins et al. 2012a) suggest the intrinsic slope varies with redshift and is bluer than that implicit in the Meurer relation.

In this work we employ a large cosmological hydrodynamic simulation ({\em MassiveBlack}-II) to make predictions for the intrinsic UV continuum slope in high-redshift star forming galaxies and contrast the attenuation inferred assuming these with those from the M99 relation. This work is organised as follows: In Section \ref{sec:r} we describe how the relationship between the inferred level of dust attenuation and the observed UV continuum slope is affected by various assumptions including the choice of intrinsic slope (\S\ref{sec:int}) and reddening law (\S\ref{sec:redlaw}). In Section \ref{sec:implications} we consider the implications of changing the these assumptions on the level of dust attenuation inferred from recent observations of the UV continuum slopes of high-redshift galaxies. Finally in Section \ref{sec:c} we present our conclusions. 

Throughout this work magnitudes are calculated using the $AB$ system (Oke \& Gunn 1983). We assume Salpeter (1955) stellar initial mass function (IMF), i.e.: $\xi(m)={\rm d}N/{\rm d}m\propto m^{-2.35}$.

\section{Relating the observed UV continuum slope to dust attenuation}\label{sec:r}

The effect of applying either an SMC-{\em like} or Calzetti et al. (2000) law to the UV continuum causes the observed slope $\beta_{\,\rm obs}$ to redden relative to the intrinsic value. Using the observed UV continuum slope as a diagnostic of dust attenuation in high-redshift star forming galaxies then essentially requires knowledge of two factors: (1) the intrinsic UV continuum slope, and (2) the reddening law. 

The relationship between the observed slope and the attenuation at a given wavelength $\lambda$ can then be defined simply as:
\begin{equation}\label{eq:abeta}
A_{\lambda}=D_{\lambda}\times[\beta_{\,\rm obs}-\beta_{\,\rm int}],
\end{equation} 
where $\beta_{\rm int}$ is the intrinsic UV continuum slope, and $\beta_{\rm obs}$ is the observed slope. The factor $D_{\lambda}$ relates a change in $\beta$ to the attenuation at $\lambda$ ($A_{\lambda}$), i.e. $D_{\lambda}={\rm d}A_{\lambda}/{\rm d}\beta$, and is sensitive to the choice of reddening law. In the following sections we discuss each of these factors in turn.
 
\subsection{Factors affecting the intrinsic UV continuum slope and predictions from numerical models of galaxy formation}\label{sec:int}

The intrinsic UV continuum slope of a stellar population is determined by the joint distribution of stellar masses, ages and metallicities. These properties are, in turn, determined by the initial mass function (IMF) together with the star formation and metal enrichment histories. The effect of each of these factors was explored in detail by Wilkins et al. (2012a).  

\subsubsection{Predictions from numerical models of galaxy formation}

By using a galaxy formation model to predict both the star formation and metal enrichment histories we can predict the intrinsic UV continuum slope of galaxies (see also Wilkins et al. 2012a). Here we make use of a state-of-the-art cosmological hydrodynamic simulation of structure formation: {\em MassiveBlack}-II. The {\em MassiveBlack}-II simulation is performed using the cosmological TreePM-Smooth Particle Hydrodynamics (SPH) code {\sc P-Gadget}, a hybrid version of the parallel code {\sc Gadget2} (Springel 2005) tailored to run on the new generation of Petaflop scale supercomputers. {\em MassiveBlack}-II includes $N_{\rm par}=2\times 1792^{3}\approx 11.5$ billion particles in a volume of $10^{6}\,{\rm Mpc}^{3}/h^{3}$ ($100\,{\rm Mpc}/h$ on a side), is run to $z=0$, and includes not only gravity and hydrodynamics but also additional physics for star formation (Springel \& Hernquist 2003), black holes and associated feedback processes (Di Matteo et al. 2008, Di Matteo et al. 2012). Spectral energy distributions (SEDs) of galaxies are generated using the stellar population synthesis (SPS) code {\sc Pegase.2} (Fioc \& Rocca-Volmerange 1997, 1999). This simulation has been demonstrated to reproduce the observed intrinsic UV luminosity function (LF) at $z=5-8$ (with the exception of the faint-end slope at $z=5$, see Wilkins et al. {\em in-prep}). 

\begin{figure}
\centering
\includegraphics[width=20pc]{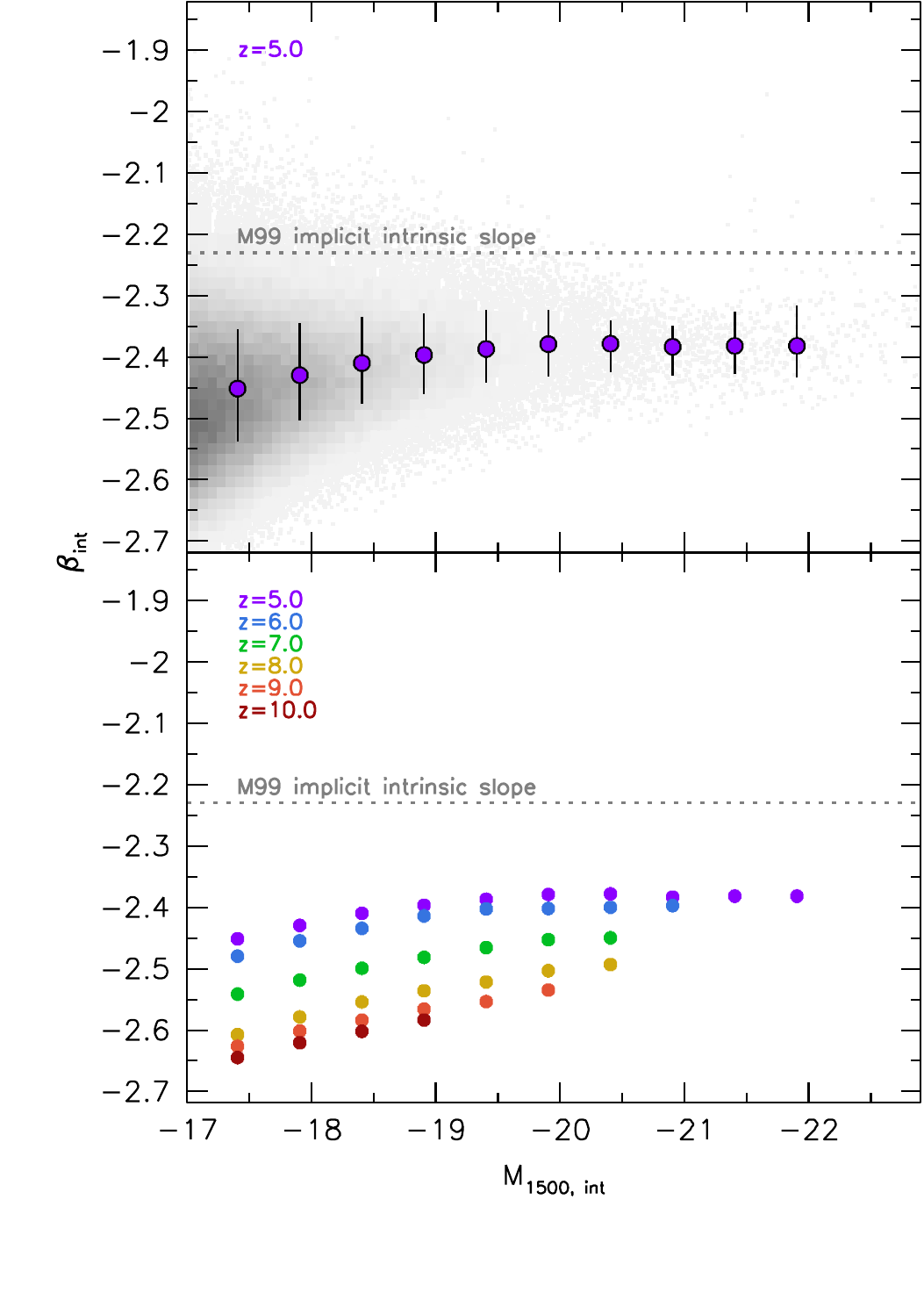}
\caption{The intrinsic UV continuum slope $\beta_{\,\rm int}$ as a function of UV luminosity predicted by the {\em MassiveBlack}-II simulation. The upper panel shows both a 2D number density histogram and the median slope (and $16^{\rm th}-84^{\rm th}$ percentile range) in several luminosity bins. The shade of the 2D density histogram denotes the number of galaxies contributing to each bin on a linear scale. Where there are fewer than 25 galaxies contributing to each bin on the 2D histogram the galaxies are plotted individually. The lower panel shows only the median slope as a function of absolute magnitude but for a range of redshifts ($z\in\{5,6,7,8,9,10\}$) highlighting the important redshift evolution. The horizontal dotted line in both panels denotes the intrinsic slope implicit in the M99 relation ($\beta=-2.23$).}
\label{fig:MB.UVC}
\end{figure}

In Figure \ref{fig:MB.UVC} the median intrinsic UV continuum slope, as a function of the intrinsic UV luminosity, for galaxies $z\in\{5,6,7,8,9,10\}$ predicted by this simulation are shown. For simplicity the UV continuum slope $\beta$ is measured using two artificial rest-frame UV bandpasses uniformly encompassing $1300-2700{\rm\AA}$. This range is similar to the range accesible for observations of galaxies at $z\sim 3-6$ using WFC3 on the {\em Hubble Space Telescope}. However, at present, for sources located at $z\sim 7$ and above a smaller range of the rest-frame UV continuum is accesible using WFC3 observations. With this in mind we also determine the UV continuum slope of galaxies at $z\sim 7$ using a shorter rest-frame UV wavelength baseline $1300-2100{\rm\AA}$ roughly corresponding to the range probed by the WFC3 $J_{f125w}$ and $H_{f160w}$ filters at $z=7$. The effect of changing to this reduced wavelength baseline reduces the intrinsic value of $\beta$ by $\sim 0.03$. For clarity Figure \ref{fig:MB.UVC} presents the UV continuum slope over the full $1300-2700{\rm\AA}$ baseline, however when we come to interpret observations at $z\sim 7$ we make use of the predicted intrinsic values of $\beta$ using the smaller baseline. This reduced wavelength baseline can also have a significant effect when determining the level of dust attenuation inferred from an observed slope through a change to the parameter $D_{\lambda}$. This is discussed in more detail in \S\ref{sec:redlaw}.   

Apparent in Figure \ref{fig:MB.UVC} is a strong trend with redshift such that galaxies at higher-redshift typically have bluer UV continuum slopes ($\beta_{\,\rm int}\approx -2.55$ at $z=8$ while $\beta_{\,\rm int}\approx -2.40$ at $z=5$). This reflects the typically younger stellar populations (and thus the presence of more high-mass stars) and lower metallicities in higher-redshift systems (Wilkins et al. {\em in prep}). There is also a trend with intrinsic UV luminosity such that more luminous objects typically have redder intrinsic UV continuum slopes. This is again a consequence of older stellar populations and higher metallicities, though this time in more luminous systems. However, this effect is fairly small with $\beta_{\,\rm int}$ increasing by less than $0.1$ as $M_{1500}=-18\to-21$ at $z=7$. These predictions closely match the luminosity and redshift trends predicted using the {\sc galform} semi-analytical galaxy formation model utilised in Wilkins et al. (2012a). However, the normalisation differs slightly due to a difference in choice of initial mass function (Wilkins et al. 2012a assumed a Kennicutt 1983 IMF).

\subsection{The choice of reddening law}\label{sec:redlaw}

The factor, $D_{\lambda}$, relating the change in $\beta$ due to the UV attenuation can be easily calculated for reddening laws in the literature. For the Calzetti et al. (2000) attenuation curve we find this factor to be $D_{1600}=2.10$. For the Large and Small Magellanic Clouds (LMC and SMC) extinction curves (using Pei et al. 1992) we obtain values of $D_{1600}=1.63$ and $D_{1600}=1.12$ respectively while for the core collapse supernovae (CCSNe) curve of Bianchi \& Schneider (2007) we find $D_{1600}=2.49$. The implication of the smaller values of $D_{\lambda}$ obtained for the SMC and LMC laws is that for the same {\em observed} UV continuum slope a smaller (and significantly smaller assuming the LMC law) UV attenuation would be inferred. These values are based on the same wavelength baseline used to measure the UV continuum slope ($1300-2700{\rm\AA}$). If the luminosity baseline is reduced to ($1300-2100{\rm\AA}$), as is appropriate when interpreting the UV continuum slope of sources observed by WFC3 at $z\sim 7$, these values become $1.84$, $0.96$, $1.47$ and $1.90$ for the Calzetti et al. (2000), SMC, LMC, and CCSNe curves respectively. The strong sensitivity of $D_{1600}$ for the CCSNe curve reflects the structure in the curve around $2000-2500{\rm\AA}$.

\subsection{The {\em Meurer} relation}

An alternative to explicitly assuming an intrinsic slope and reddening law is to use an empirical relation based on observations of the UV continuum slope and some observational measurement of the UV attenuation (typically from far-IR observations). A prominent and widely used example of such an empirical calibration is the Meurer et al. (1999) relation. This relation is defined as:
\begin{equation}\label{eq:M99}
A_{1600}=4.43+1.99\times\beta_{\,\rm obs}.
\end{equation} 
Re-arranging this relation to match the form of Eqn. \ref{eq:abeta} reveals the M99 relation implicitly assumes an intrinsic slope of $\beta_{\,\rm obs}=-2.23$ and factor relating the change in $\beta$ to UV attenuation of $D_{1600}=1.99$. The value of $D_{1600}$ is unsurprisingly similar to that derived assuming the Calzetti et al. (2000) reddening law. The implicit intrinsic slope ($\beta_{\,\rm obs}=-2.23$) however is significantly redder than that predicted for star forming galaxies at high-redshift. Considering the similar values of $D_{1600}$ this then suggests assuming the intrinsic slope(s) predicted by {\em MassiveBlack}-II combined with the Calzetti et al. (2000) reddening law would yield larger UV attenuations for the same observed value of $\beta$.

\section{Implications for inferring dust attenuation at high-redshift}\label{sec:implications}

In the previous section we described how the observed UV continuum slope is sensitive to the choice of intrinsic slope and reddening law. We now investigate how these different assumptions affect the interpretation of observations of the UV continuum slope at high-redshift.

\subsection{The observed UV continuum slope}

\begin{figure}
\centering
\includegraphics[width=20pc]{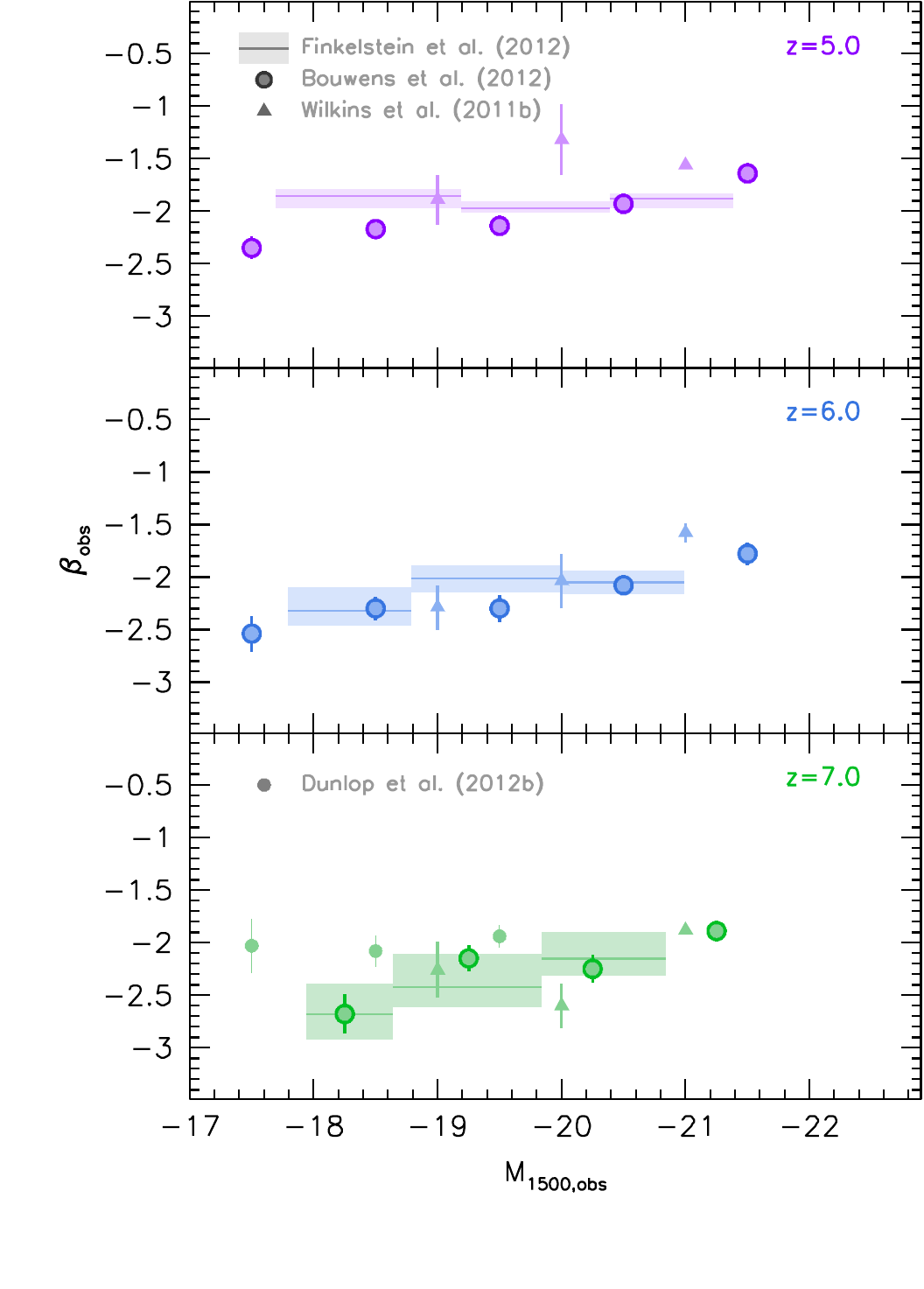}
\caption{Measurements of the observed UV continuum slope in the literature at $z\in\{5,6,7\}$. For the measurements from Dunlop et al. (2012b) we utilise their determination of $\beta$ using a power-law methodology (i.e. column 4 of table 1 of Dunlop et al. 2012b). Finkelstein et al. (2012a) present their result in three luminosity bins referenced on the local value of $L^{*}$ ($L<0.25\times L^{*}$, $0.25\times L^{*}<L<0.75\times L^{*}$, $L>0.75\times L^{*}$). The median is represented by the solid line while the shaded regions show the uncertainties. For the $L<0.25\times L^{*}$ and $L>0.75\times L^{*}$ bins the minimum and maximum bounds respectively are chosen to reflect the approximate range of contributing galaxy luminosities.}
\label{fig:observations}
\end{figure}

As noted in the introduction numerous groups have recently studied the observed UV continuum slopes of high-redshift star forming (Lyman-break selected\footnote{While several studies use a photometric redshift selection technique most of the significance comes from the strong Lyman-$\alpha$ break feature thus they are {\em de facto} LBGs.}) galaxies (e.g. Stanway et al. 2005, Bouwens et al. 2009, Bunker et al. 2010, Bouwens et al. 2010a, Finkelstein et al. 2010, Wilkins et al. 2011a, Dunlop et al. 2012a, Finkelstein et al. 2012a, Bouwens et al. 2012, Rogers et al. 2012, Dunlop et al. 2012b). A compilation of measurements of $\beta$ from some of these studies (at $z=5-7$) is shown in Figure \ref{fig:observations}. 

The measurements from different groups are largely consistent, with typical observed slopes of $\beta\approx -2$. There are however continuing conflicting claims regarding a correlation with redshift and/or luminosity. For example, Bouwens et al. (2009), Wilkins et al. (2011b), Bouwens et al. (2012), and Finkelstein et al. (2012) all find evidence of a correlation with redshift (such that galaxies at lower redshift have redder observed slopes) while Dunlop et al. (2012a) do not. Bouwens et al. (2009), Bouwens et al. (2010), Wilkins et al. (2011b), and Bouwens et al. (2012) find a significant correlation with luminosity while Dunlop et al. (2012a) and Dunlop et al. (2012b) do not. Finkelstein et al. (2012a) do not observe a strong relation between the observed $\beta$ and UV luminosity at $z\sim 4$ or $5$ in their full sample\footnote{However, Finkelstein et al. (2012a) note if they restrict the faintest luminosity bin to include only galaxies from the deeper HUDF dataset they do find evidence for a correlation with luminosity.} though they do note a trend at $z\sim 7$ (though this may be the result of selection bias). Several possible explanations, relating to both the methodology used to measure $\beta$, the UV luminosity\footnote{Finkelstein et al. (2012a) do find a correlation with stellar mass, this is found both observationally and predicted by {\em MassiveBlack-II} to correlate with UV luminosity.}, and the bias in the sample are suggested to reconcile this apparent discrepancy in both Bouwens et al. (2012) and Finkelstein et al. (2012a).

\subsection{The effect of changing assumptions on the inferred level of dust attenuation}

While there is some disagreement between observational studies in this paper we are simply interested in exploring the effect of different assumptions on the inferred level of dust attenuation. With this in mind we consider only the Bouwens et al. (2012) observations in our subsequent analysis, though note it would be equally valid to use the Finkelstein et al. (2012a) results considering the similar sample size and luminosity baseline.  

We then consider four sets of assumptions controlling the conversion of the observed UV continuum slope to a UV attenuation: (1) the M99 relation, (2) the intrinsic slope(s) predicted by {\em MassiveBlack}-II combined with a Calzetti et al. (2000) reddening law (C00+MBII scenario), (3) the intrinsic slope(s) predicted by {\em MassiveBlack}-II combined with a Pei et al. (1992) empirical SMC reddening law (SMC+MBII), and (4) the intrinsic slope(s) predicted by {\em MassiveBlack}-II combined with collapse supernovae (CCSNe) curve of Bianchi \& Schneider (2007) (CCSNe+MBII). 

\begin{figure}
\centering
\includegraphics[width=20pc]{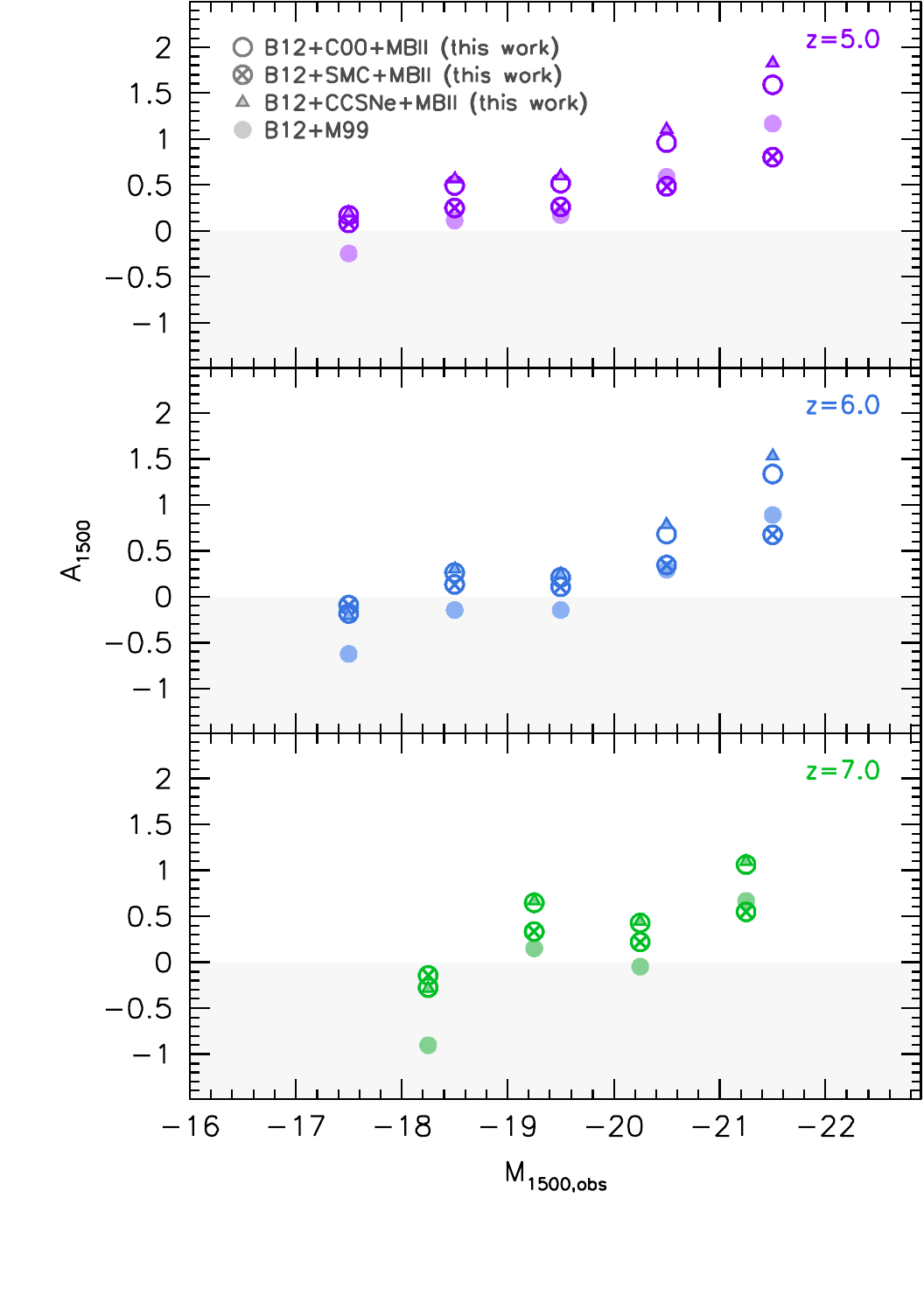}
\caption{UV attenuation inferred from observations (Bouwens et al. 2012: B12) of the UV continuum slope at $z\in\{5,6,7\}$ using three different assumptions: (1) The empirically determined M99 relation, (2) intrinsic UV continuum slopes predicted from {\em MassiveBlack}-II and a Calzetti et al. (2000) reddening law, (3) intrinsic UV continuum slopes predicted from {\em MassiveBlack}-II and a SMC type reddening law, and (4) intrinsic UV continuum slopes predicted from {\em MassiveBlack}-II and the CCSNe reddening law of Bianchi \& Schneider (2007).}
\label{fig:attenuation_all}
\end{figure}

\begin{figure}
\centering
\includegraphics[width=20pc]{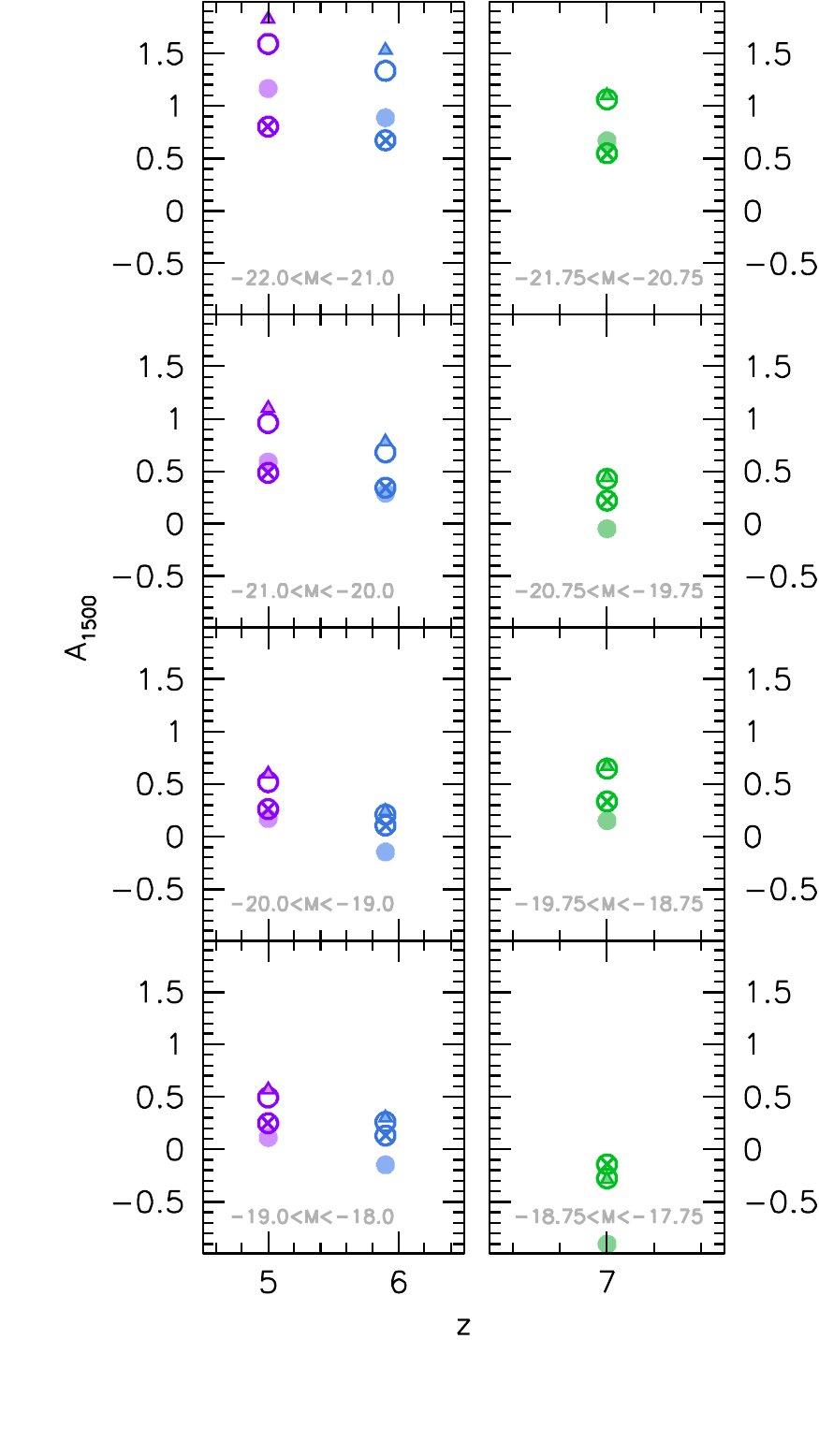}
\caption{Similar to Figure \ref{fig:attenuation_all} but instead showing the UV attenuation as a function of redshift (for several luminosity bins). Note: the luminosity bins utilised for the $z\sim 7$ observations is different from those at $z\sim 5$ and $z\sim 6$.}
\label{fig:attenuation_all_z}
\end{figure}

The level of (UV) attenuation ($A_{1500}$) inferred from the Bouwens et al. (2012) observations using these four sets of assumptions are shown in both Figure \ref{fig:attenuation_all} (where we show the inferred attenuation as a function of observed UV luminosity for the three redshifts bins) and Figure \ref{fig:attenuation_all_z} (where we show the inferred attenuation as a function redshift for several luminosity bins). As noted previously for the $z\sim 7$ sample we utilise the intrinsic UV continuum and values of $D_{1500}$ measured assuming a smaller wavelength baseline given the coverage of the WFC3 observations.

The difference in the inferred attenuation ($A_{1500}$) between the M99 relation and C00+MBII scenario is largely fixed as a function of luminosity reflecting the similar functional form of the Calzetti et al. (2000) reddening curve and that implicit in the M99 relation (i.e. the value of $D_{\lambda}$). This difference does however evolve with redshift due to evolution of the intrinsic slope predicted from the simulations (and the non-evolution of the intrinsic slope implicit in the M99 relation). At $z\approx 5$ the difference in $A_{1500}$ is $\approx 0.35\,{\rm mag}$ and increases to $\approx 0.5\,{\rm mag}$ at $z\approx 7$. Given that the application of the M99 relation to observations at $z\approx 7$ results in zero (or even negative) attenuations at fainter luminosities this difference has a strong impact. This increase in the inferred attenuation brings the level of attenuation even for the faintest luminosities consistent with zero. The effect of assuming a CCSNe curve is also similar to the C00+MBII scenario considering the similar values of $D_{1500}$. 

The difference between assuming the M99 relation and the SMC+MBII scenario is more nuanced. While the value of $D_{1500}$ is {\em larger} for the Calzetti law than the SMC law (i.e. for a similar increase in $\beta$ a greater attenuation is inferred using the Calzetti law) the intrinsic slope predicted by MBII is {\em bluer} than that implicit in the M99 relation. For very-blue observed values of $\beta$ ($<-2$) this difference in the implicit intrinsic slope is the most important factor and results in a greater level of attenuation inferred assuming the SMC+MBII scenario than utilising M99. A consequence of this is that at low luminosities the level of dust inferred assuming the SMC+MBII scenario does not drastically fall below zero. At redder observed values of $\beta$ ($>-2$) this trend reverses and the difference in $D_{1500}$ dominates. Thus at very-high luminosities the level of attenuation inferred assuming the SMC+MBII scenario is lower than simply assuming M99 (though this is only true for the most luminous point at $z=5-6$).

As can be seen in Figure \ref{fig:attenuation_all_z} the adoption of a decreasing intrinsic UV continuum slope to higher redshift (as opposed to the fixed value implicit in the M99 relation) also weakens, though does not destroy, the inferred (decreasing) trend of UV attenuation with redshift.  

\section{Conclusions}\label{sec:c}

It has been established, at low-redshift, that the UV continuum slope is a useful diagnostic of dust attenuation in typical star forming galaxies. However, the relation between the level of attenuation and the observed slope is sensitive to both the assumed intrinsic slope and reddening law.

While it is possible to use an empirically determined calibration (such as the Meurer et al. 1999 relation) this nevertheless implicitly assumes an intrinsic UV continuum slope (in this case $\beta_{\,\rm int}=-2.23$) which may not be valid for very-high redshift where the UV continuum slope is the only accesible diagnostic of dust attenuation. Indeed, we find that results from the large cosmological hydrodynamical simulation {\em MassiveBlack}-II suggest the intrinsic slope both varies strongly with redshift, and, at $z>4$ is bluer than the value implicit in the Meurer relation. 

Making use of the intrinsic UV continuum slopes predicted by {\em MassiveBlack}-II (and the Calzetti et al. 2000 reddening law) we re-interpret recent observations of the UV continuum slope finding levels of dust attenuation $0.35-0.5\,{\rm mag}$ (where the difference is sensitive to the redshift and luminosity) higher than simply applying the Meurer relation. If instead we utilise an SMC-{\em like} redding law (combined with the intrinsic slope predicted by {\em MassiveBlack}-II) we find levels of attenuation similar to those inferred using the Meurer relation with the exception of the lowest luminosities (where we infer a higher level of attenuation).

\subsection*{Acknowledgements}

We would like to thank the anonymous referee for their detailed and extremely useful report which greatly improved this manuscript. SMW and AB acknowledge support from the Science and Technology Facilities Council. WRC acknowledges support from an Institute of Physics/Nuffield Foundation funded summer internship at the University of Oxford. RACC thanks the Leverhulme Trust for their award of a Visiting Professorship at the University of Oxford. The simulations were run on the Cray XT5 supercomputer Kraken at the National Institute for Computational Sciences. This research has been funded by the National Science Foundation (NSF) PetaApps program, OCI-0749212 and by NSF AST-1009781.

\bsp

\end{document}